\documentclass[aps,pra,reprint,groupedaddress]{revtex4-2}

\usepackage{amssymb}
\usepackage{amsmath}
\usepackage{graphicx}

\begin{document}

\title{Quantum Computation by Cooling}

\author{Jaeyoon Cho}
\email{choooir@gmail.com}
\affiliation{Department of Physics and Research Institute of Natural Science, Gyeongsang National University, Jinju 52828, Korea}
\date{\today}
\begin{abstract}
Adiabatic quantum computation is a paradigmatic model aiming to solve a computational problem by finding the many-body ground state encapsulating the solution. However, its use of an adiabatic evolution depending on the spectral gap of an intricate many-body Hamiltonian makes its analysis daunting. While it is plausible to directly cool the final gapped system of the adiabatic evolution instead, the analysis of such a scheme on a general ground is missing. Here, we propose a specific Hamiltonian model for this purpose. The scheme is inspired by cavity cooling, involving the emulation of a zero-temperature reservoir. Repeated discarding of ancilla reservoir qubits extracts the entropy of the system, driving the system toward its ground state. At the same time, the measurement of the discarded qubits hints at the energy level structure of the system as a return. We show that quantum computation based on this cooling procedure is equivalent in its computational power to the one based on quantum circuits. We then exemplify the scheme with a few illustrative use cases for combinatorial optimization problems. In the first example, the cooling is free from any local energy minima, reducing the scheme to Grover's search algorithm with a few improvements. In the second example, the cooling suffers from abundant local energy minima. To circumvent this, we implant a mechanism in the Hamiltonian so that the population trapped in the local minima can tunnel out by high-order transitions. We support this idea with a numerical simulation for a particular combinatorial optimization problem. We also discuss its application to preparing quantum many-body ground states, arguing that the spectral gap is a crucial factor in determining the time scale of the cooling.
\end{abstract}

\maketitle

\section{Introduction}

Identifying the ground states of quantum many-body Hamiltonians and understanding their characteristics are fundamental in physics. 
The ground states encode the spectral properties of the Hamiltonian~\cite{has06,nac06,has07,ara13,bra13,cho14,cho18} and represent the essential quantumness of the system.
The latter becomes particularly evident in gapped systems, in which the ground state predominantly determines the entire properties at sufficiently low temperature.
This lays the foundation for various exotic quantum phenomena, such as quantum Hall effects and superconductivity~\cite{wen07}.

Computing many-body ground states is extremely hard in general:
for classical Hamiltonians, it falls into the NP-hard complexity class, and for quantum Hamiltonians, it is classified as QMA-hard~\cite{kit02,kem03,kem06}.
The difficulty arises mainly from the geometric frustration, i.e., the stark mismatch between local and global energy minima. The problem is combinatorial in nature:
one needs to find the optimal configurations of data among exponentially many, mutually conflicting ones.
This is indeed the core challenge of combinatorial optimization problems~\cite{kor18}.
It is well-established that combinatorial optimization problems can be reduced to the task of finding the ground states of Ising-like spin models~\cite{bar82,mez02,luc14}.

It is widely believed that quantum computers are not universally efficient in solving general ground-state problems, even for classical Hamiltonians.
Nonetheless, we still need to find out specific classes of such problems that benefit from quantum computation and are yet practically useful.
Also, there's a need to devise diverse quantum algorithms for them, each having its unique advantages.
We have a number of tools available for such purposes~\cite{kad98,bha22,far01,aha08,alb18,zho20,far14,had19,kan17,cer21,til22,ver09,cho11,cub23} (Ref.~\cite{cub23} contains a nice summary of earlier schemes).
The most prominent is adiabatic quantum computation (AQC)~\cite{far01,aha08,alb18}.
The AQC exploits the adiabatic evolution of the instantaneous ground state when the Hamiltonian is varied.
The time scale of the operation is thus governed by the minimal spectral gap during the evolution.
However, analyzing the spectral gap is extremely difficult, leaving the performance of AQC largely unknown.
There are also hybrid approaches, such as quantum approximate optimization algorithms (QAOA)~\cite{far14,had19,zho20} and variational quantum eigensolvers (VQE)~\cite{kan17,cer21,til22}, aided by classical optimization subroutines.

As mentioned, the difficulty in analyzing AQC arises as the initial and final Hamiltonians intermingle.
On the other hand, the properties of the final Hamiltonian alone can be largely unveiled, even though its ground state remains unknown.
Given this, it is natural to consider schemes to directly cool the final Hamiltonian system to its ground state.
While earlier schemes based on dissipative engineering~\cite{ver09,cho11,cub23} seem to achieve this goal, a closer inspection reveals that they actually lack desirable properties as a cooling method.
To clarify this, it is instructive to discuss classical and quantum Hamiltonians separately.
By the former, we mean those Hamiltonians diagonalized in the computational basis.

For classical Hamiltonians, the earlier dissipative schemes are reduced to a classical random search.
There are a few fundamental reasons.
They implement a completely-positive trace-preserving (CPTP) map that transforms the state $\rho$ to $\sum_k E_k \rho E_k^\dagger$~\cite{nie11}.
Here, each Kraus operator $E_k$ directly accesses the associated {\em local} part of the system, driving it into a certain direction.
However, as the energy of the whole system is a global quantity, determining the optimal direction based solely on a small part is not possible (note that this was the very origin of the geometric frustration, rendering the problem intractable).
For this reason, the state is randomized when the measurement projects the state into an undesired subspace.
Note, however, that this entire process is essentially equivalent to a mere random guess.
Furthermore, for classical Hamiltonians, they do not involve any operations in non-computational bases.
This means that they are, in fact, identical to classical probabilistic models.
In the case of quantum Hamiltonians, the problem is more involving: one needs to find an appropriate basis containing the ground state.
This is the place where the above CPTP map comes into play.
Apart from the role of fixing the basis, however, the above-mentioned limitations persist.

Simply put, both cooling and heating coexist in the above-mentioned schemes.
A genuine cooling scheme should rely on quantum transitions that take place exclusively in the direction of lowering energy.
It is important to understand that a heating-prevention mechanism requires an external agent, which globally accesses the system, gathers its energy, and relaxes it into a Markovian reservoir.
This process is exactly provided by cavity cooling methods~\cite{cho15}.
This paper aims at materializing this concept as a general quantum computation model and discuss its efficiency in some illustrative scenarios.
Our model can be paralleled with AQC for their similarity in motivation.
On the other hand, unlike AQC, our model exploits a non-unitary process.
For brevity, we will refer to our model as cooling-based quantum computation (CQC) hereafter.

Being a dissipative process, CQC inherits the intrinsic advantages of the earlier dissipative schemes, alleviating the demand of precise system controls to some extent.
On top of this, CQC takes true quantum advantages.
In fact, we show that CQC is equivalent to quantum circuit models in terms of computational power.
This argument is similar to the established one regarding the computational equivalence between AQC and quantum circuit models~\cite{aha08}.
All three frameworks are thus computationally equivalent.
However, this assertion needs to be interpreted with care.
For instance, the equivalence of AQC to CQC is established by translating the unitary evolution of AQC into a quantum circuit, which is then converted into a Hamiltonian for CQC.
However, this resulting Hamiltonian differs from the original problem Hamiltonian of the AQC.
It is an open question whether the feasibility of ground-state preparation by AQC implies that of the same Hamiltonian by CQC, and vice versa.

Any ground-state preparation scheme, including AQC, inevitably suffers from the issue of local energy minima~\cite{mez02,ami08,wer23}.
In CQC, one can design tunneling transitions out of local minima if the information is given on their nature.
This flexibility also constitutes the advantage of CQC.
In practice, one could try various transition terms in the Hamiltonian, while observing the occurrence of the transition by monitoring cavity photons.
This, in turn, provides a method to inspect the energy-level structure of the Hamiltonian.
For example, the presence or absence of a finite spectral gap above the ground state could be inspected for unsolvable Hamiltonians.

In this work, much of our focus will be placed on CQC for combinatorial optimization problems, which have favorable features for CQC and allow for the analytical treatment.
We consider two extreme cases.
The first case is free from local energy minima, but the transition rate is superpolynomially small in the system size.
This case turns out to be equivalent to Grover's quantum search algorithm, aside from some advantages that CQC brings.
The second case is an opposite limit, wherein the transition rate is maximized, but local energy minima begin to pose challenges.
The local minima can be overcome by introducing high-order tunnel-out transitions.
We perform the numerical simulation for a particular combinatorial optimization problem to support this idea.
Finally, we briefly discuss the case of quantum Hamiltonians.
This case is heavily system-dependent and more complicated to analyze.
We argue that the spectral gap above the ground state is a crucial factor in determining the time scale of the cooling.
This is a natural consequence of the energy-time uncertainty principle.

%%%%
%%%%
\section{Notation and working principles}

Consider an $N$-qubit system described by Hamiltonian
\begin{equation}
    H_P = \sum_{z=0}^{2^N-1} E(z) |z\rangle_P \langle z|,
    \label{eq:problemh}
\end{equation}
where $E(z) \le E(z')$ for $z < z'$.
Let us call $H_P$ a problem Hamiltonian.
We aim to find the ground state $|0\rangle_P$.
Transitions among the energy levels $|z\rangle_P$ are allowed by introducing in the Hamiltonian what we call a transition term $H_T$, which does not commute with $H_P$.
$H_T$ should be designed to suit the needs of the given problem.
We will consider two simple examples below.
Here, we assume the energy level spacings of $H_P$ are characterized by an energy scale $\Delta>0$, and so is the norm $\| H_T \|$.
Our aim is to introduce a parameter $0 < \lambda \ll \Delta$ to treat $\lambda H_T$ as a perturbation to $H_P$.

If we consider the transition problem described by Hamiltonian $H_P + \lambda H_T$, the transition from $|z\rangle_P$ to $|z'\rangle_P$ is allowed only when $|E(z)-E(z')| \lesssim | {}_P\langle z' | \lambda H_T | z \rangle_P |$, where the influence of the degeneracy is ignored.
In order to ensure $E(z') < E(z)$, we introduce cavity field $c_m$ with resonant frequency $\omega_m \gg | {}_P\langle z' | \lambda H_T | z \rangle_P |$ and couple it to the transition.
Here, the subscript $m$ is the index for different cavities.
Let us denote by $|\cdot\rangle_m$ the photon number state of the cavity.
If the cavity is initially empty, the transition from $|z\rangle_P |0\rangle_m$ to $|z'\rangle_P |1\rangle_m$ is allowed only when $E(z') \simeq E(z) - \omega_m$, providing a mechanism to prevent heating.

Introducing $M$ different cavity modes, our final Hamiltonian reads
\begin{equation}
  H = H_P + \sum_{m=1}^{M} \omega_m c_m^\dagger c_m + \lambda H_T \otimes \left[
      a_0 I + \sum_{m=1}^{M}(c_m + c_m^\dagger)
  \right].
  \label{eq:h}
\end{equation}
Here, $a_0 \in \{0, 1\}$ determines whether the transitions preserving the cavity photon numbers are allowed.
This term helps the state escape from local minima through high-order transitions.
In practice, it is generally enough to restrict the maximum number of photons in each cavity to one.
In this case, $M$ cavity modes can be simulated with $M$ qubits.

Our cooling protocol is a stochastic process performed by \emph{repeating} the following cycle.
To simplify the description, suppose $a_0=0$ and the system is initially in state $|z_0\rangle_P \bigotimes_{m=1}^M |0\rangle_m $ with arbitrary $z_0$.
The energy of this state is $E(z_0)$.
First, we let the system evolve unitarily under Hamiltonian~\eqref{eq:h}.
This causes the system to undergo coherent oscillations between the initial state and a superposition of all states $|z_m\rangle_P |1\rangle_m$ satisfying $|E(z_0) - \{ E(z_m) + \omega_m \}| \lesssim | {}_P\langle z_m | \lambda H_T | z_0 \rangle_P|$, where it is understood that $|z_m\rangle_P$ may represent a superposition of multiple energy levels of $H_P$ in cases of degeneracy.
After a certain period (explained below), each cavity state is measured.
If a cavity photon, say $|1\rangle_m$, is detected, the system collapses into state $|z_m\rangle_P |1\rangle_m$, where $|z_m\rangle_P $ has a lower energy than $|z_0\rangle_P$ by $\omega_m$.
In this case, we empty the cavity by resetting the state to $|0\rangle_m$ so that the next cooling cycle starts from the lower-energy state $|z_m\rangle_P \bigotimes_{m=1}^M |0\rangle_m$.
If no cavity photon is detected, the system simply collapses back into the initial state of the current cycle.
Note that in any case, the cycle ends by emptying all cavities.
This \emph{information-discarding} process simulates a zero-temperature reservoir.
During the cycle, heating is avoided as long as $\omega_m \gg | {}_P\langle z_m | \lambda H_T | z_0 \rangle_P|$ for all $m$.
Precise control of the operation time is not essential as the cooling cycle is repeated without the risk of heating; given that the probability of detecting a cavity photon in a single cycle is finite (depending on the duration), the expected number of cycles to lower the energy is also finite.
Having said that, the optimal duration of the unitary evolution is given by the inverse of the oscillation frequency multiplied by $\pi/2$.
However, as this optimal time varies for each transition path and multiple oscillations take place simultaneously, finding a single optimal duration is generally challenging.
The case for $a_0=1$ is elaborated in Sec.~\ref{sec:comb}.

The measurement of the cavity state allows us to monitor the progress of cooling.
It is a reasonable strategy to repeat the cooling cycle until cavity photons are not detected for a sufficient number of cycles.
This signals with high probability that the system has reached either the ground state or a local minimum.
For NP problems that are guaranteed to have at least one solution, distinguishing between the two cases is straightforward.
Otherwise, to increase confidence, one may repeat the computation, possibly varying the transition term $H_T$.
Given that an unknown ground state is not verifiable in general, the decision is probabilistic after all.

We will implicitly assume that the evolution by the Hamiltonian~\eqref{eq:h} is simulated on a universal quantum computer, which is reasonable given the complexity of the Hamiltonian.
In our analysis below, the polynomial overhead associated with this implementation is not taken into account.

%%%%
%%%%
\section{Equivalence to the quantum circuit model}
%%%%
%%%%

\begin{figure}
    \centerline{\includegraphics[width=\columnwidth]{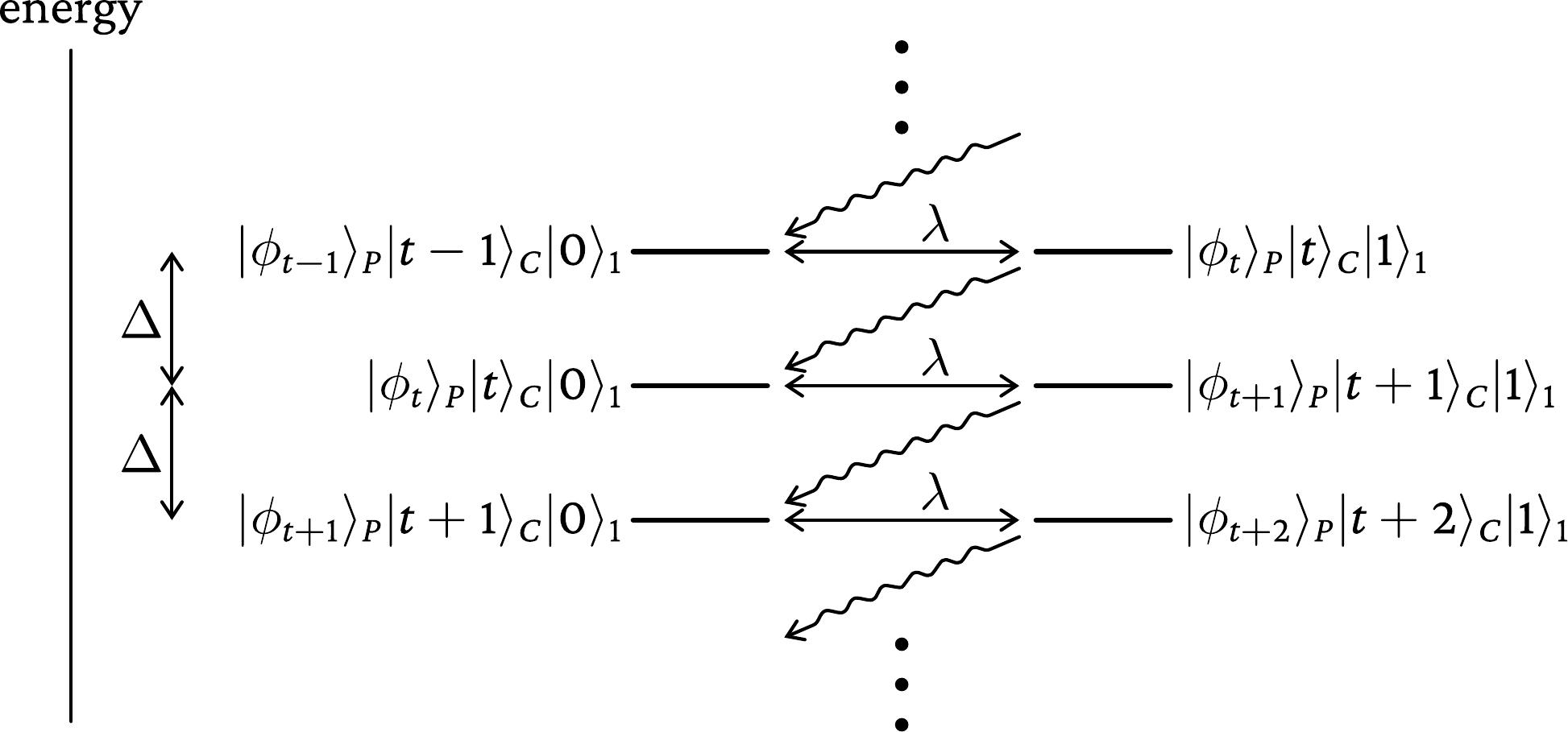}}
    \caption{The transition diagram into which a quantum circuit is converted for CQC. $|\phi_t\rangle_P$, $|t\rangle_C$, and $|\cdot\rangle_1$ denotes, respectively, the state of the quantum circuit at the $t$-th time step, the clock state to label the time step, and the number state of the cavity. $\Delta$ is the energy difference between the adjacent levels and $\lambda$ is the Rabi frequency of the designated transition. The wavy lines represent the transition after detecting and clearing the cavity photon.}
    \label{fig:transition}
\end{figure}

Ref.~\cite{aha08} shows that AQC is equivalent to the quantum circuit model in computational power up to polynomial overhead.
The forward direction of the equivalence is straightforward because the entire procedure of AQC is a unitary transformation, which can be simulated efficiently in quantum circuits.
Showing the other direction relies on the clock state $|t\rangle_C$ tagged to the state at the $t$-th time step of the quantum-circuit computation.

A similar equivalence relation can be shown for CQC.
Again, showing that CQC can be simulated efficiently in a quantum circuit is straightforward.
We thus focus on the other direction: a quantum-circuit computation can be efficiently simulated in CQC.
Suppose a quantum algorithm is run on a quantum circuit in $T$ time steps, where $U_t^P$ is applied to the qubits at the $t$-th time step.
Let $|\phi_t\rangle_P \equiv U_t^P U_{t-1}^P\cdots U_1^P |\phi_0\rangle_P$ be the state right after applying $U_t^P$, and tag this state with the clock state as $|\phi_t\rangle_P |t\rangle_C$.
The idea is to get $|\phi_t\rangle_P |t\rangle_C$ transformed into $|\phi_{t+1}\rangle_P |t+1\rangle_C$ in the course of an energy-lowering transition.
This is enabled by choosing the terms in Hamiltonian~\eqref{eq:h} as
\begin{eqnarray}
  H_P &=& - \sum_{t=0}^{T} t \Delta  |t \rangle_C \langle t|,\\
  H_T &=& \sum_{t=0}^{T-1} (U_{t+1}^P\otimes |t+1 \rangle_C \langle t| + \mathrm{H.c.}),
\end{eqnarray}
where the parameter $\Delta>0$ is an arbitrary energy scale,
and the other parmeters are chosen as $M=1$, $\omega_1 = \Delta$, and $\alpha_0 = 0$.
If the initial state is chosen to be $|\phi_0\rangle_P  |0\rangle_C  |0\rangle_1$, the entire transition dynamics is restricted to a subspace spanned by $\{ |\phi_t\rangle_P |t\rangle_C, 0\le t\le T \}$, as depicted in Fig.~\ref{fig:transition}.
Note that the off-diagonal element ${}_{P,C}\langle \phi_{t+1},t+1| \lambda H_T | \phi_t, t\rangle_{P,C} = \lambda$ is constant for every $t$.

The cooling occurs in the same manner as the sideband cooling~\cite{esc03}.
Initially, the system undergoes the Rabi oscillation between $|\phi_0\rangle_P |0\rangle_C |0\rangle_1$ and $|\phi_1\rangle_P |1\rangle_C |1\rangle_1$ with Rabi frequency $\lambda$.
After a certain duration, the cavity state is measured.
In case of measuring $|1\rangle_1$, we reset the cavity state to $|0\rangle_1$.
The subsequent evolution then becomes a Rabi oscillation between $|\phi_1\rangle_P |1\rangle_C |0\rangle_1$ and $|\phi_2\rangle_P |2\rangle_C |1\rangle_1$.
If $|0\rangle_1$ is measured, the same Rabi oscillation simply restarts.
Repeating this cycle thus makes the state cascade down the energy levels.
The optimal duration of the Rabi oscillation is $\pi/2\lambda$, for which the cavity photon is detected with unit probability.
However, precise timing is not essential, as mentioned in the previous section.
If the cavity photon is detected $n$ times, the current state of the system is $|\phi_n\rangle_P$.
Consequently, the average time required to reach the final state is polynomial in $T$.
As the Rabi frequency of the oscillation $\lambda \ll \Delta$ is constant at every cycle, the computation time is $\mathcal{O}(T/\Delta)$.

The above Rabi oscillation relies on the conventional rotating wave approximation, which neglects terms that do not preserve the energy of the unperturbed Hamiltonian $H_P$.
The omitted term in this approximation induces heating albeit with a small probability.
However, this heating is essentially harmless because it does not take the state out of the subspace spanned by $\{ |\phi_t\rangle_P |t\rangle_C\}$.
Even if the heating occurs (with a small probability), the mere effect is to regress the cooling by a single step, causing the transition $|\phi_t\rangle_P |t\rangle_C \rightarrow |\phi_{t-1}\rangle_P |t-1\rangle_C$.
Consequently, the overall computation time of $\mathcal{O}(T/\Delta)$ remains rigorously correct.

As mentioned in the introduction, the equivalence of CQC to the quantum circuit model should be interpreted carefully.
What we have shown above is that computations in one model can be converted to those in the other model up to polynomial overhead, indicating that they are computationally equivalent.
However, it is important to note that the two models are motivated differently.
CQC is basically a scheme to prepare a many-body ground state.
While CQC is advantageous in cases where finding a ground state naturally accomplishes a computation, such as in combinatorial optimization, or finding a ground state itself is the objective, there is no inherent reason to prefer CQC over the quantum circuit model for ordinary quantum algorithms.
The same argument applies to AQC as it shared the same motivation with CQC~\cite{aha08}.

%%%%
%%%%
\section{Combinatorial optimization by cooling-based quantum computation\label{sec:comb}}
%%%%
%%%%

%%%%
%%%%
\subsection{Notation}
%%%%
%%%%

Consider combinatorial optimization problems that aim to find the optimal solutions from a set of $2^N$ possible configurations represented by $N$-bit string $z \equiv z_1 z_2 \cdots z_N$.
The optimality of each configuration is evaluated by a non-negative cost function $E(z)$.
The objective is to identify configurations $z$ that minimize $E(z)$.
This problem is straightforwardly translated into the ground-state finding problem, where the cost function $E(z)$ is identified with the energy eigenvalue in Hamiltonian~\eqref{eq:problemh}.
Here, we set $E(z)$ as integer multiples of a parameter $\Delta>0$.
For example, for boolean satisfiability (SAT) problems, $E(z)$ is the number of unsatisfied clauses by the configuration $z$~\cite{kor18}.
The cavity frequencies in Hamiltonian~\eqref{eq:h} can then be chosen as integer multiples of $\Delta$.
By choosing $M$ to be the maximum of $E(z)$, all possible transitions in the problem Hamiltonian can be coupled to a cavity transition.

%%%%
%%%%
\subsection{Reproduction of Grover's search algorithm}
%%%%
%%%%

Grover's quantum search algorithm is paradigmatic in quantum information theory~\cite{gro96,nie11}.
It can be recast as an algorithm to find the zeros of a binary function $f(z)\in\{0,1\}$ from the set of $2^N$ configurations $z$.
Provided that $f(z)$ is computed only by an external agent, called the oracle, Grover's algorithm requires $\mathcal{O}(2^{N/2})$ oracle calls to find the solution, while the number becomes $\mathcal{O}(2^N)$ for the best classical algorithm.
The quadratic speedup of Grover's algorithm is proven to be optimal unless the structure of the function $f(z)$ is exploited somehow~\cite{ben97,boy98,zal99}.

To perform an analogous task in CQC, we adjust the Hamiltonian~\eqref{eq:h} as
\begin{equation}
  H = \sum_{z=0}^{2^N-1} \Delta f(z) |z\rangle_P \langle z| + \Delta c_1^\dagger c_1 + \lambda \bigotimes_{i=1}^{N} \frac{I_i + X_i}{2} \otimes (c_1 + c_1^\dagger),
  \label{eq:grover}
\end{equation}
where $I_i$ and $X_i$ denote the identity and Pauli $X$ operators acting on qubit $i$, respectively.
Let $n_0$ be the number of zeros of $f(z)$.
For the initial state $2^{-N/2} \sum_z |z\rangle_P |0\rangle_1$, the problem Hamiltonian $H_P$ effectively becomes two-dimensional, where the Hilbert space is spanned by
\begin{equation}
    |\phi_0\rangle_P = \frac{1}{\sqrt{n_0}} \sum_{f(z)= 0} |z\rangle_P
\end{equation}
and
\begin{equation}
    |\phi_1\rangle_P = \frac{1}{\sqrt{2^N - n_0}} \sum_{f(z) \not= 0} |z\rangle_P.
\end{equation}
The cooling process is thus very simple.
The unitary evolution of the cooling cycle is a Rabi oscillation between $|\phi_0\rangle_P |0\rangle_1$ and $|\phi_1\rangle_P |1\rangle_1$ occurring with the Rabi frequency ${}_P\langle \phi_0 | \lambda \bigotimes_i \frac{I_i + X_i}{2} | \phi_1 \rangle_P \simeq \lambda n_0^{1/2} 2^{-N/2}$.
By repeating the cooling cycle until the cavity photon is detected, one can obtain the solution state $|\phi_0\rangle_P$.
As the optimal duration of the Rabi oscillation is proportional to the inverse of the Rabi frequency, the computation time is given by $\mathcal{O}(\sqrt{2^N/n_0})$, identical to that of Grover's algorithm.
In comparison to Grover's algorithm, one advantage of the CQC implementation is that the end of computation is signaled by the detection of the cavity photon.
This is especially beneficial when the number of solutions $n_0$ is unknown.

In this example, the problem Hamiltonian has only two energy levels corresponding to the solution state and the rest.
A naturally following question is if the performance can be improved when the problem Hamiltonian has more than two energy levels.
It can be seen that with the transition term chosen above, more energy levels just make the computation slower.
To see this, note that the quadratic speedup comes from the collective transition, the rate of which increases with the degeneracy of the energy level.
The increase in the number of energy levels leads to a decrease in the degeneracy of each level, which diminishes the collective effect.
Moreover, it results in an increased number of transitions required to reach the ground state.

%%%%
%%%%
\subsection{Alternative approach}
%%%%
%%%%

In the previous example reproducing Grover's search algorithm, the transition term generates an all-to-all interaction with the identical transition strength.
This makes the problem Hamiltonian utterly featureless except for each energy level having a different degeneracy.
While the absence of any local energy minimum is advantageous, the transition strength decreases superpolynomially with the number of qubits.

In this subsection, we consider an opposite limit, where the transition term is given by
\begin{equation}
    H_T = \sum_{i=1}^{N} X_i.
    \label{eq:x}
\end{equation}
The off-diagonal element ${}_P\langle z | H_T | z' \rangle_P$ is then non-vanishing and equals one, independently of $N$, only when the Hamming distance between $z$ and $z'$ is one.
While the transition strength is maximum, local energy minima now become an issue.
As the distance between two configurations is at most $N$, any state $| z \rangle_P$ can reach the ground state in at most $N$ transitions in principle.
However, we do not know a general rule for every state to take the right path.
For this reason, we take the direction of lowering the energy, even though it does not necessarily coincide with the direction of approaching the ground state.
This mismatch is the origin of local energy minima.

Note that every state $|z\rangle_P$ is linked to $N$ different states by transition.
It is instructive to envisage this as an $N$-regular graph
with $2^N$ vertices.
Each vertex is endowed with a potential determined by the problem Hamiltonian, and the transition occurs along the edges toward the direction of not increasing the potential.
Once the population is trapped in a local potential minimum, a high-order transition is needed to get out of it.

\begin{figure*}
    \centerline{\includegraphics[width=0.9\textwidth]{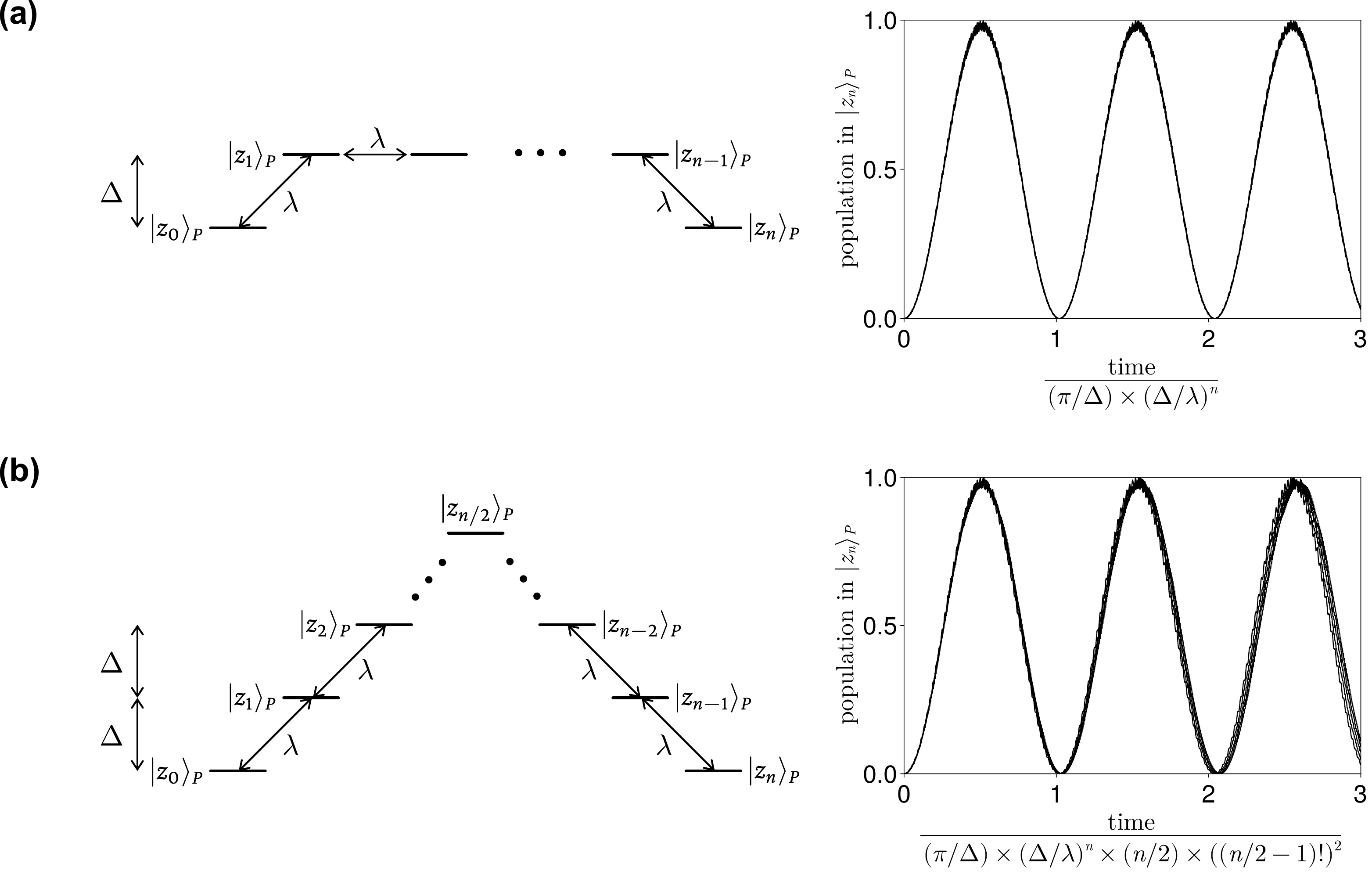}}
    \caption{Demonstration of representative high-order transitions for $\Delta = 10\lambda$, where $\Delta$ and $\lambda$ denote the energy gap and the Rabi frequency, respectively. The plots on the right show the population in $|z_n\rangle_P$ as a function of the rescaled time, starting from the initial state $|z_0\rangle_P$. The results for (a) $n=2,3,4,5,6$ and (b) $n=2,4,6,8,10$ are overlaid in the plots. The indistinguishability of the individual curves confirms the reliability of the perturbation treatment.}
    \label{fig:high-order}
\end{figure*}

To elucidate this mechanism, consider an $n$-th order transition through the sequence $|z_0\rangle_P \leftrightarrow |z_1\rangle_P \leftrightarrow \cdots \leftrightarrow |z_n\rangle_P$.
For the moment, suppose that $\alpha_0=1$ in Hamiltonian~\eqref{eq:h} and there is no cavity mode involved.
For such a transition to occur, the first requirement is $E(z_0) = E(z_n)$.
In this case, the $n$-th order perturbation theory yields the effective transition rate $\mathcal{O}[\lambda (\lambda/\Delta)^{n-1}]$.
However, if the Stark shifts of $|z_0\rangle_P$ and $|z_n\rangle_P$ differ, the effective detuning breaks the condition $E(z_0) = E(z_n)$, suppressing the transition.
Consequently, the most prominent transitions result from cases where the energy differences satisfy the condition $E(z_{i+1}) - E(z_i) = E(z_{n-(i+1)}) - E(z_{n-i})$ for all $i$ (see Fig.~\ref{fig:high-order} for typical cases).

Among various high-order transition channels, we demonstrate two representative cases in Fig.~\ref{fig:high-order}.
The Hamiltonian employed for the figure is
\begin{equation}
    H = \sum_{j=0}^{n}E(z_j) |z_j\rangle_P \langle z_j| + \lambda \sum_{j=0}^{n-1} \left(|z_j\rangle_P \langle z_{j+1} | + |z_{j+1}\rangle_P \langle z_{j} | \right).
\end{equation}
In Fig.~\ref{fig:high-order}(a), we have chosen $E(z_0)=E(z_n)=0$, $E(z_1)=E(z_2)= \cdots = E(z_{n-1}) = \Delta$, and $\lambda = \Delta / 10$.
In this case, the $n$-th-order perturbation theory states that the oscillation between $|z_0\rangle_P$ and $|z_n\rangle_P$ occurs with frequency $\Omega_n \simeq \Delta (\frac{\lambda}{\Delta})^n$.
The plot on the right in Fig.~\ref{fig:high-order}(a) shows the population in $|z_n\rangle_P$ with respect to the \emph{rescaled} time (in unit of $\pi / \Omega_n$) when the initial state is $|z_0\rangle_P$.
The plot includes the curves for $n \in \{2,3,4,5,6\}$ and they are almost indistinguishable, confirming the reliability of the perturbation treatment.
In Fig.~\ref{fig:high-order}(b), we have chosen $E(z_j) = j \Delta$ for $j \le n/2$ and $E(z_j) = (n-j) \Delta$ for $j \ge n/2$.
In this case, the oscillation frequency is given by $\Omega_n \simeq \Delta (\frac{\lambda}{\Delta}\frac{\lambda}{2\Delta} \cdots \frac{\lambda}{(n/2-1)\Delta})^2 \frac{\lambda}{(n/2)\Delta}$ from the perturbation theory.
The plot on the right shows the population in $|z_n\rangle_P$ for the initial state $|z_0\rangle_P$.
The curves for $n\in \{2, 4, 6, 8, 10\}$ are plotted with respect to the rescaled time and they are again almost indistinguishable.

Now, let us take the cavity modes into account.
Note that the time evolution remains identical when any intermediate transition $|z_i\rangle_P \leftrightarrow |z_{i+1}\rangle_P$ is replaced by $|z_i\rangle_P |0\rangle_m \leftrightarrow |z_{i+1}'\rangle_P |1\rangle_m$ with $E(z_{i+1}) = E(z_{i+1}') + \omega_m$.
This is always possible provided $E(z_{i+1}') < E(z_{i+1})$ and the appropriate cavity mode exists.
By incorporating multiple cavity modes, various tunnel-out transition channels can appear.
For instance, the lambda-type-like transitions characterized by $E(z_1) - E(z_0) = E(z_{n-1}) - E(z_n) + \omega_m$ with $n \ge 2$, equivalent to the case in Fig.~\ref{fig:high-order}(a), are expected to play an important role in overcoming local minima.

To illustrate this approach to the cooling, we consider a particular integer factoring algorithm (not to be confused with Shor's algorithm~\cite{sho94,nie11}).
Specifically, we follow the integer multiplication procedure as in the elementary arithmetic and turn it into a combinatorial optimization problem.
The problem Hamiltonian is constructed as follows.
Consider a multiplication of two 3-bit integers $\bar{x} \equiv (x_2 x_1 x_0)_2$ and $\bar{y} \equiv (y_2 y_1 y_0)_2$ resulting in a 6-bit integer $\bar{z} \equiv (z_5 z_4 \cdots z_0)_2$, incorporating additional four carry bits $\bar{c} \equiv (c_3 c_2 c_1 c_0)_2$.
For a given value of $\bar{z}$, we turn each calculation step into an energy term in the problem Hamiltonian.
Here, 10 qubits are needed to encode $\bar{x}$, $\bar{y}$, and $\bar{c}$, while $\bar{z}$ is hard-coded in the problem Hamiltonian.
For example, the first energy term has a value zero if $x_0 y_0 = z_0$ and $\Delta$ otherwise, the second has a value zero if $x_1 y_0 + x_0 y_1 = (c_0 z_1)_2$ and $\Delta$ otherwise, and so on.
The ground state, representing the solution of the factoring, has energy zero as it satisfies all the conditions, and has a two-fold degeneracy as $\bar{x}$ and $\bar{y}$ are interchangeable.

\begin{figure}
    \centerline{\includegraphics[width=\columnwidth]{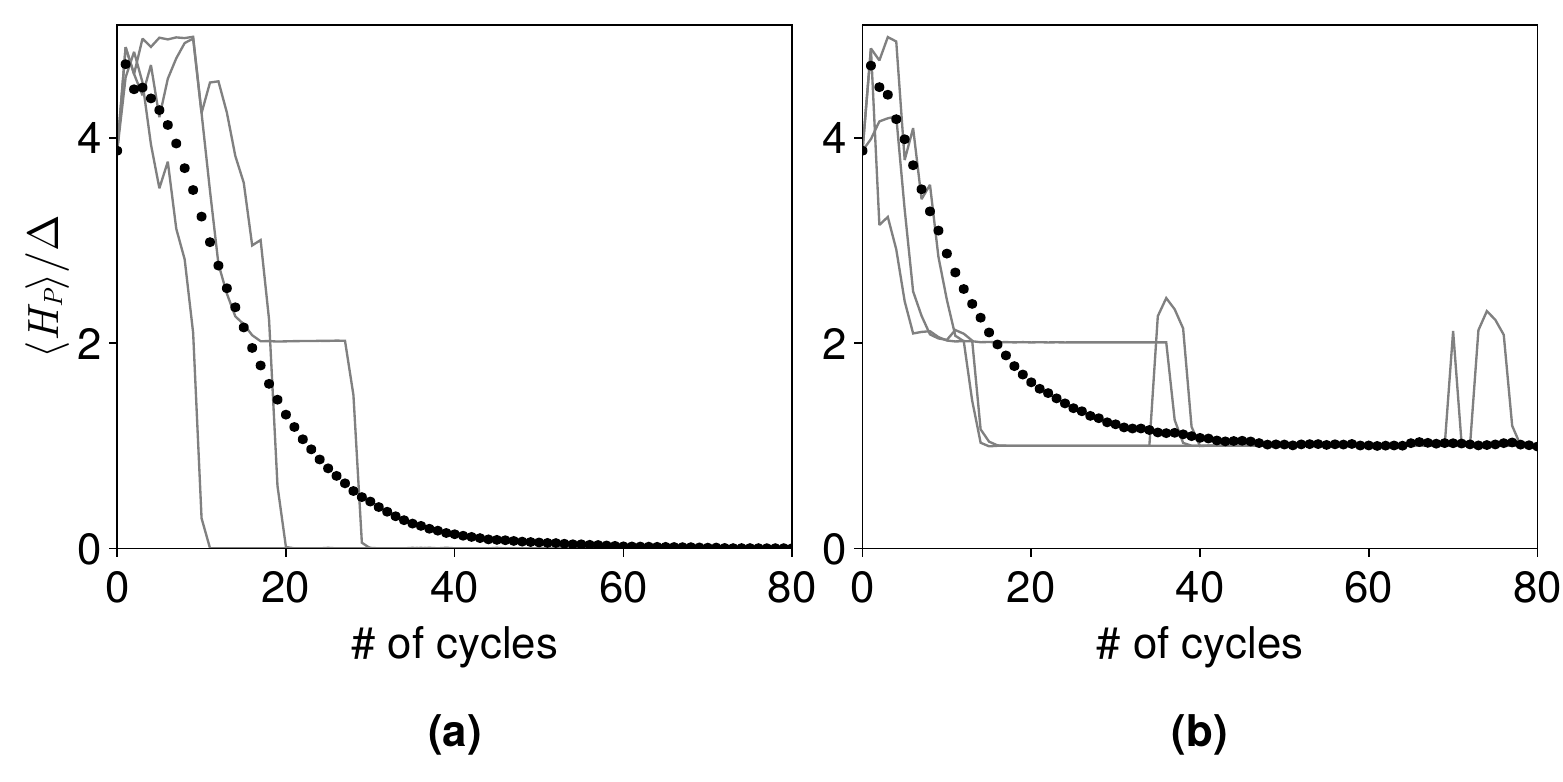}}
    \caption{Numerical simulation of CQC for integer factoring of 35. Dotted curves represent the energy with respect to the number of cooling cycles, averaged over $10^3$ samples, for (a) $\alpha_0 = 1$ and (b) $\alpha_0 = 0$. Gray curves represent typical individual trajectories.}
    \label{fig:factoring}
\end{figure}

Fig.~\ref{fig:factoring} shows the results of our numerical simulation for $\bar{z} = 35$.
We take the transition term as in Eq.~\eqref{eq:x} and set the parameters in Hamiltonian~\eqref{eq:h} as $M = 3$, $\omega_m = m \Delta$, $\lambda = \Delta / 10$, and $a_0 = 1$.
The duration of a single cooling cycle is chosen to be $\pi / 2\lambda$.
The simulation is performed by exact diagonalization.
Fig.~\ref{fig:factoring}(a) shows that the energy of the system
vanishes in time, indicating that the system evolves
into the ground state.
As a comparison, we have changed $a_0$ to zero for Fig.~\ref{fig:factoring}(b) to test the effectiveness of the mechanism for tunneling out from local minima.
Without such a mechanism, the system almost always evolves into local minima, which appear to be present at the first excited level.

A remark is in order.
Our choice of the parameter $\lambda=\Delta/10$, chosen to fulfill the requirement that the transition rate is much smaller than the energy difference, appears effective, as demonstrated in Fig.~\ref{fig:factoring}(a).
However, it should be noted that the optimal choice of $\lambda$ depends on the characteristics and size of the problem.
In general, smaller $\lambda$ is preferable as the collective effect enhances the transition rate.
This effect is particularly important for combinatorial optimization problems, which are characterized by large degeneracies in the problem Hamiltonian.
As the degeneracies are generally very large in the middle of the energy spectrum, the collective effect is also stronger in that region.
Consequently, there exists a parametric regime of $\lambda$ where the heating is sufficiently avoided for the ground state, but not for higher energy states.
On the other hand, by choosing smaller $\lambda$, leading to slower transitions, one could reach the regime where heating is avoided throughout the entire energy spectrum.
While the cooling is feasible in both cases, it is not clear which case is more efficient.

%%%%
%%%%
\section{Discussion}
%%%%
%%%%

In this work, we have introduced the notion of cooling-based quantum computation along with a few illustrative examples.
An important part is to choose an appropriate transition term, which generates off-diagonal elements in the diagonal basis of the problem Hamiltonian.
There exists a trade-off between the connectivity and strength of the transition.
We have discussed two opposite limits when the ground state is the solution of a combinatorial optimization problem.
In both cases, the running time generally increases superpolynomially with the system size.
It is an open question whether the performance can be substantially improved by interpolating between the two opposites.

The idea of CQC could be used to prepare the ground states of general \emph{gapped} quantum Hamiltonians.
A qualitative argument is the following.
Recall that the transition from $|z_1\rangle_P$ to $|z_2\rangle_P$ occurs when the condition $|E(z_2) - \{E(z_1) - \omega_m\}| \lesssim |{}_P \langle z_2 | \lambda H_T | z_1 \rangle_P|$ is satisfied, where a cavity mode with frequency $\omega_m$ is involved.
The key point is that a single cavity mode enables transitions with the associated energy difference up to a margin determined by the spectral \emph{linewidth} $|{}_P \langle z_2 | \lambda H_T | z_1 \rangle_P|$.
To ensure that the transition can distinguish the ground state from the excited state, this linewidth should be sufficiently smaller than the spectral gap above the ground state.
Consequently, a finite spectral gap sets the upper bound on the transition rate, hence the time scale of the cooling.
Crucially, a finite spectral gap in turn allows to have a finite linewidth for transitions.
The entire spectral range of the problem Hamiltonian, which is $\mathcal{O}(N)$, can then be divided into $\mathcal{O}(N/\lambda)$ sectors.
This means that a polynomial number of cavity modes suffice to cover the entire spectrum, provided that one can design appropriate transition terms $H_T$ along with the cavity modes to establish transition channels to the ground state for all the energy levels.
Of course, the practical application of this idea would heavily depend on the specific Hamiltonian and hence analytically challenging.
The requirement of a finite spectral gap is a natural consequence of the energy-time uncertainty principle.

An interesting open question is the precise relationship between CQC and AQC.
In particular, the abundance and depth of local energy minima in CQC might be strongly correlated with the spectral gap during AQC~\cite{mez02,ami08,wer23}.
If not, the two methods can differ in efficiency for particular ground-state findings.
In fact, the transition terms $H_T$ in Eqs.~\eqref{eq:grover} and \eqref{eq:x} are the two conventional initial Hamiltonians in AQC.
Consequently, when Trotter-decomposed, the infinitesimal unitary evolutions of CQC with Eqs.~\eqref{eq:grover} and \eqref{eq:x} are identical to those of AQC, although their arrangements and the inclusion or exclusion of cavities make differences.
This might be a clue to understanding one of the two more deeply from inspecting the other.

\begin{acknowledgments}
This work was supported by the National Research Foundation (NRF) of Korea under Grant No.~NRF-2022R1A4A1030660.
\end{acknowledgments}

%\bibliographystyle{apsrev4-2}
%\bibliography{references}

\begin{thebibliography}{39}%
\makeatletter
\providecommand \@ifxundefined [1]{%
 \@ifx{#1\undefined}
}%
\providecommand \@ifnum [1]{%
 \ifnum #1\expandafter \@firstoftwo
 \else \expandafter \@secondoftwo
 \fi
}%
\providecommand \@ifx [1]{%
 \ifx #1\expandafter \@firstoftwo
 \else \expandafter \@secondoftwo
 \fi
}%
\providecommand \natexlab [1]{#1}%
\providecommand \enquote  [1]{``#1''}%
\providecommand \bibnamefont  [1]{#1}%
\providecommand \bibfnamefont [1]{#1}%
\providecommand \citenamefont [1]{#1}%
\providecommand \href@noop [0]{\@secondoftwo}%
\providecommand \href [0]{\begingroup \@sanitize@url \@href}%
\providecommand \@href[1]{\@@startlink{#1}\@@href}%
\providecommand \@@href[1]{\endgroup#1\@@endlink}%
\providecommand \@sanitize@url [0]{\catcode `\\12\catcode `\$12\catcode `\&12\catcode `\#12\catcode `\^12\catcode `\_12\catcode `\%12\relax}%
\providecommand \@@startlink[1]{}%
\providecommand \@@endlink[0]{}%
\providecommand \url  [0]{\begingroup\@sanitize@url \@url }%
\providecommand \@url [1]{\endgroup\@href {#1}{\urlprefix }}%
\providecommand \urlprefix  [0]{URL }%
\providecommand \Eprint [0]{\href }%
\providecommand \doibase [0]{https://doi.org/}%
\providecommand \selectlanguage [0]{\@gobble}%
\providecommand \bibinfo  [0]{\@secondoftwo}%
\providecommand \bibfield  [0]{\@secondoftwo}%
\providecommand \translation [1]{[#1]}%
\providecommand \BibitemOpen [0]{}%
\providecommand \bibitemStop [0]{}%
\providecommand \bibitemNoStop [0]{.\EOS\space}%
\providecommand \EOS [0]{\spacefactor3000\relax}%
\providecommand \BibitemShut  [1]{\csname bibitem#1\endcsname}%
\let\auto@bib@innerbib\@empty
%</preamble>
\bibitem [{\citenamefont {Hastings}\ and\ \citenamefont {Koma}(2006)}]{has06}%
  \BibitemOpen
  \bibfield  {author} {\bibinfo {author} {\bibfnamefont {M.~B.}\ \bibnamefont {Hastings}}\ and\ \bibinfo {author} {\bibfnamefont {T.}~\bibnamefont {Koma}},\ }\href {https://doi.org/10.1007/s00220-006-0030-4} {\bibfield  {journal} {\bibinfo  {journal} {Communications in Mathematical Physics}\ }\textbf {\bibinfo {volume} {265}},\ \bibinfo {pages} {781} (\bibinfo {year} {2006})}\BibitemShut {NoStop}%
\bibitem [{\citenamefont {Nachtergaele}\ and\ \citenamefont {Sims}(2006)}]{nac06}%
  \BibitemOpen
  \bibfield  {author} {\bibinfo {author} {\bibfnamefont {B.}~\bibnamefont {Nachtergaele}}\ and\ \bibinfo {author} {\bibfnamefont {R.}~\bibnamefont {Sims}},\ }\href {https://doi.org/10.1007/s00220-006-1556-1} {\bibfield  {journal} {\bibinfo  {journal} {Communications in Mathematical Physics}\ }\textbf {\bibinfo {volume} {265}},\ \bibinfo {pages} {119} (\bibinfo {year} {2006})}\BibitemShut {NoStop}%
\bibitem [{\citenamefont {Hastings}(2007)}]{has07}%
  \BibitemOpen
  \bibfield  {author} {\bibinfo {author} {\bibfnamefont {M.~B.}\ \bibnamefont {Hastings}},\ }\href {https://doi.org/10.1088/1742-5468/2007/08/P08024} {\bibfield  {journal} {\bibinfo  {journal} {Journal of Statistical Mechanics: Theory and Experiment}\ }\textbf {\bibinfo {volume} {2007}},\ \bibinfo {pages} {P08024} (\bibinfo {year} {2007})}\BibitemShut {NoStop}%
\bibitem [{\citenamefont {Arad}\ \emph {et~al.}()\citenamefont {Arad}, \citenamefont {Kitaev}, \citenamefont {Landau},\ and\ \citenamefont {Vazirani}}]{ara13}%
  \BibitemOpen
  \bibfield  {author} {\bibinfo {author} {\bibfnamefont {I.}~\bibnamefont {Arad}}, \bibinfo {author} {\bibfnamefont {A.}~\bibnamefont {Kitaev}}, \bibinfo {author} {\bibfnamefont {Z.}~\bibnamefont {Landau}},\ and\ \bibinfo {author} {\bibfnamefont {U.}~\bibnamefont {Vazirani}},\ }\href {https://doi.org/https://doi.org/10.48550/arXiv.1301.1162} {\bibinfo {title} {An area law and sub-exponential algorithm for {1D} systems}},\ \bibinfo {note} {arXiv:1301.1162}\BibitemShut {NoStop}%
\bibitem [{\citenamefont {Brand\~ao}\ and\ \citenamefont {Horodecki}(2013)}]{bra13}%
  \BibitemOpen
  \bibfield  {author} {\bibinfo {author} {\bibfnamefont {F.~G. S.~L.}\ \bibnamefont {Brand\~ao}}\ and\ \bibinfo {author} {\bibfnamefont {M.}~\bibnamefont {Horodecki}},\ }\href {https://doi.org/10.1038/nphys2747} {\bibfield  {journal} {\bibinfo  {journal} {Nature Physics}\ }\textbf {\bibinfo {volume} {9}},\ \bibinfo {pages} {721} (\bibinfo {year} {2013})}\BibitemShut {NoStop}%
\bibitem [{\citenamefont {Cho}(2014)}]{cho14}%
  \BibitemOpen
  \bibfield  {author} {\bibinfo {author} {\bibfnamefont {J.}~\bibnamefont {Cho}},\ }\href {https://doi.org/10.1103/PhysRevLett.113.197204} {\bibfield  {journal} {\bibinfo  {journal} {Physical Review Letters}\ }\textbf {\bibinfo {volume} {113}},\ \bibinfo {pages} {197204} (\bibinfo {year} {2014})}\BibitemShut {NoStop}%
\bibitem [{\citenamefont {Cho}(2018)}]{cho18}%
  \BibitemOpen
  \bibfield  {author} {\bibinfo {author} {\bibfnamefont {J.}~\bibnamefont {Cho}},\ }\href {https://doi.org/10.1103/PhysRevX.8.031009} {\bibfield  {journal} {\bibinfo  {journal} {Physical Review X}\ }\textbf {\bibinfo {volume} {8}},\ \bibinfo {pages} {031009} (\bibinfo {year} {2018})}\BibitemShut {NoStop}%
\bibitem [{\citenamefont {Wen}(2007)}]{wen07}%
  \BibitemOpen
  \bibfield  {author} {\bibinfo {author} {\bibfnamefont {X.-G.}\ \bibnamefont {Wen}},\ }\href@noop {} {\emph {\bibinfo {title} {Quantum {Field} {Theory} of {Many}-body {Systems}: {From} the {Origin} of {Sound} to an {Origin} of {Light} and {Electrons}}}}\ (\bibinfo  {publisher} {Oxford University Press},\ \bibinfo {address} {Oxford},\ \bibinfo {year} {2007})\BibitemShut {NoStop}%
\bibitem [{\citenamefont {Kitaev}\ \emph {et~al.}(2002)\citenamefont {Kitaev}, \citenamefont {Shen},\ and\ \citenamefont {Vyalyi}}]{kit02}%
  \BibitemOpen
  \bibfield  {author} {\bibinfo {author} {\bibfnamefont {A.~Y.}\ \bibnamefont {Kitaev}}, \bibinfo {author} {\bibfnamefont {A.}~\bibnamefont {Shen}},\ and\ \bibinfo {author} {\bibfnamefont {M.~N.}\ \bibnamefont {Vyalyi}},\ }\href@noop {} {\emph {\bibinfo {title} {Classical and {Quantum} {Computation}}}}\ (\bibinfo  {publisher} {American Mathematical Soc.},\ \bibinfo {year} {2002})\BibitemShut {NoStop}%
\bibitem [{\citenamefont {Kempe}\ and\ \citenamefont {Regev}(2003)}]{kem03}%
  \BibitemOpen
  \bibfield  {author} {\bibinfo {author} {\bibfnamefont {J.}~\bibnamefont {Kempe}}\ and\ \bibinfo {author} {\bibfnamefont {O.}~\bibnamefont {Regev}},\ }\href@noop {} {\bibfield  {journal} {\bibinfo  {journal} {Quantum Information \& Computation}\ }\textbf {\bibinfo {volume} {3}},\ \bibinfo {pages} {258} (\bibinfo {year} {2003})}\BibitemShut {NoStop}%
\bibitem [{\citenamefont {Kempe}\ \emph {et~al.}(2006)\citenamefont {Kempe}, \citenamefont {Kitaev},\ and\ \citenamefont {Regev}}]{kem06}%
  \BibitemOpen
  \bibfield  {author} {\bibinfo {author} {\bibfnamefont {J.}~\bibnamefont {Kempe}}, \bibinfo {author} {\bibfnamefont {A.}~\bibnamefont {Kitaev}},\ and\ \bibinfo {author} {\bibfnamefont {O.}~\bibnamefont {Regev}},\ }\href {https://doi.org/10.1137/S0097539704445226} {\bibfield  {journal} {\bibinfo  {journal} {SIAM Journal on Computing}\ }\textbf {\bibinfo {volume} {35}},\ \bibinfo {pages} {1070} (\bibinfo {year} {2006})}\BibitemShut {NoStop}%
\bibitem [{\citenamefont {Korte}\ and\ \citenamefont {Vygen}(2018)}]{kor18}%
  \BibitemOpen
  \bibfield  {author} {\bibinfo {author} {\bibfnamefont {B.}~\bibnamefont {Korte}}\ and\ \bibinfo {author} {\bibfnamefont {J.}~\bibnamefont {Vygen}},\ }\href {https://doi.org/10.1007/978-3-662-56039-6} {\emph {\bibinfo {title} {Combinatorial {Optimization}: {Theory} and {Algorithms}}}},\ \bibinfo {series} {Algorithms and {Combinatorics}}, Vol.~\bibinfo {volume} {21}\ (\bibinfo  {publisher} {Springer Berlin Heidelberg},\ \bibinfo {address} {Berlin, Heidelberg},\ \bibinfo {year} {2018})\BibitemShut {NoStop}%
\bibitem [{\citenamefont {Barahona}(1982)}]{bar82}%
  \BibitemOpen
  \bibfield  {author} {\bibinfo {author} {\bibfnamefont {F.}~\bibnamefont {Barahona}},\ }\href {https://doi.org/10.1088/0305-4470/15/10/028} {\bibfield  {journal} {\bibinfo  {journal} {Journal of Physics A: Mathematical and General}\ }\textbf {\bibinfo {volume} {15}},\ \bibinfo {pages} {3241} (\bibinfo {year} {1982})}\BibitemShut {NoStop}%
\bibitem [{\citenamefont {M\'ezard}\ and\ \citenamefont {Zecchina}(2002)}]{mez02}%
  \BibitemOpen
  \bibfield  {author} {\bibinfo {author} {\bibfnamefont {M.}~\bibnamefont {M\'ezard}}\ and\ \bibinfo {author} {\bibfnamefont {R.}~\bibnamefont {Zecchina}},\ }\href {https://doi.org/10.1103/PhysRevE.66.056126} {\bibfield  {journal} {\bibinfo  {journal} {Physical Review E}\ }\textbf {\bibinfo {volume} {66}},\ \bibinfo {pages} {056126} (\bibinfo {year} {2002})}\BibitemShut {NoStop}%
\bibitem [{\citenamefont {Lucas}(2014)}]{luc14}%
  \BibitemOpen
  \bibfield  {author} {\bibinfo {author} {\bibfnamefont {A.}~\bibnamefont {Lucas}},\ }\href@noop {} {\bibfield  {journal} {\bibinfo  {journal} {Frontiers in Physics}\ }\textbf {\bibinfo {volume} {2}},\ \bibinfo {pages} {5} (\bibinfo {year} {2014})}\BibitemShut {NoStop}%
\bibitem [{\citenamefont {Kadowaki}\ and\ \citenamefont {Nishimori}(1998)}]{kad98}%
  \BibitemOpen
  \bibfield  {author} {\bibinfo {author} {\bibfnamefont {T.}~\bibnamefont {Kadowaki}}\ and\ \bibinfo {author} {\bibfnamefont {H.}~\bibnamefont {Nishimori}},\ }\href {https://doi.org/10.1103/PhysRevE.58.5355} {\bibfield  {journal} {\bibinfo  {journal} {Physical Review E}\ }\textbf {\bibinfo {volume} {58}},\ \bibinfo {pages} {5355} (\bibinfo {year} {1998})}\BibitemShut {NoStop}%
\bibitem [{\citenamefont {Bharti}\ \emph {et~al.}(2022)\citenamefont {Bharti}, \citenamefont {Cervera-Lierta}, \citenamefont {Kyaw}, \citenamefont {Haug}, \citenamefont {Alperin-Lea}, \citenamefont {Anand}, \citenamefont {Degroote}, \citenamefont {Heimonen}, \citenamefont {Kottmann}, \citenamefont {Menke}, \citenamefont {Mok}, \citenamefont {Sim}, \citenamefont {Kwek},\ and\ \citenamefont {Aspuru-Guzik}}]{bha22}%
  \BibitemOpen
  \bibfield  {author} {\bibinfo {author} {\bibfnamefont {K.}~\bibnamefont {Bharti}}, \bibinfo {author} {\bibfnamefont {A.}~\bibnamefont {Cervera-Lierta}}, \bibinfo {author} {\bibfnamefont {T.~H.}\ \bibnamefont {Kyaw}}, \bibinfo {author} {\bibfnamefont {T.}~\bibnamefont {Haug}}, \bibinfo {author} {\bibfnamefont {S.}~\bibnamefont {Alperin-Lea}}, \bibinfo {author} {\bibfnamefont {A.}~\bibnamefont {Anand}}, \bibinfo {author} {\bibfnamefont {M.}~\bibnamefont {Degroote}}, \bibinfo {author} {\bibfnamefont {H.}~\bibnamefont {Heimonen}}, \bibinfo {author} {\bibfnamefont {J.~S.}\ \bibnamefont {Kottmann}}, \bibinfo {author} {\bibfnamefont {T.}~\bibnamefont {Menke}}, \bibinfo {author} {\bibfnamefont {W.-K.}\ \bibnamefont {Mok}}, \bibinfo {author} {\bibfnamefont {S.}~\bibnamefont {Sim}}, \bibinfo {author} {\bibfnamefont {L.-C.}\ \bibnamefont {Kwek}},\ and\ \bibinfo {author} {\bibfnamefont {A.}~\bibnamefont {Aspuru-Guzik}},\ }\href {https://doi.org/10.1103/RevModPhys.94.015004} {\bibfield  {journal} {\bibinfo  {journal} {Reviews of Modern Physics}\ }\textbf {\bibinfo {volume} {94}},\ \bibinfo {pages} {015004} (\bibinfo {year} {2022})}\BibitemShut {NoStop}%
\bibitem [{\citenamefont {Farhi}\ \emph {et~al.}(2001)\citenamefont {Farhi}, \citenamefont {Goldstone}, \citenamefont {Gutmann}, \citenamefont {Lapan}, \citenamefont {Lundgren},\ and\ \citenamefont {Preda}}]{far01}%
  \BibitemOpen
  \bibfield  {author} {\bibinfo {author} {\bibfnamefont {E.}~\bibnamefont {Farhi}}, \bibinfo {author} {\bibfnamefont {J.}~\bibnamefont {Goldstone}}, \bibinfo {author} {\bibfnamefont {S.}~\bibnamefont {Gutmann}}, \bibinfo {author} {\bibfnamefont {J.}~\bibnamefont {Lapan}}, \bibinfo {author} {\bibfnamefont {A.}~\bibnamefont {Lundgren}},\ and\ \bibinfo {author} {\bibfnamefont {D.}~\bibnamefont {Preda}},\ }\href {https://doi.org/10.1126/science.1057726} {\bibfield  {journal} {\bibinfo  {journal} {Science}\ }\textbf {\bibinfo {volume} {292}},\ \bibinfo {pages} {472} (\bibinfo {year} {2001})}\BibitemShut {NoStop}%
\bibitem [{\citenamefont {Aharonov}\ \emph {et~al.}(2008)\citenamefont {Aharonov}, \citenamefont {van Dam}, \citenamefont {Kempe}, \citenamefont {Landau}, \citenamefont {Lloyd},\ and\ \citenamefont {Regev}}]{aha08}%
  \BibitemOpen
  \bibfield  {author} {\bibinfo {author} {\bibfnamefont {D.}~\bibnamefont {Aharonov}}, \bibinfo {author} {\bibfnamefont {W.}~\bibnamefont {van Dam}}, \bibinfo {author} {\bibfnamefont {J.}~\bibnamefont {Kempe}}, \bibinfo {author} {\bibfnamefont {Z.}~\bibnamefont {Landau}}, \bibinfo {author} {\bibfnamefont {S.}~\bibnamefont {Lloyd}},\ and\ \bibinfo {author} {\bibfnamefont {O.}~\bibnamefont {Regev}},\ }\href@noop {} {\bibfield  {journal} {\bibinfo  {journal} {SIAM Review}\ }\textbf {\bibinfo {volume} {50}},\ \bibinfo {pages} {755} (\bibinfo {year} {2008})}\BibitemShut {NoStop}%
\bibitem [{\citenamefont {Albash}\ and\ \citenamefont {Lidar}(2018)}]{alb18}%
  \BibitemOpen
  \bibfield  {author} {\bibinfo {author} {\bibfnamefont {T.}~\bibnamefont {Albash}}\ and\ \bibinfo {author} {\bibfnamefont {D.~A.}\ \bibnamefont {Lidar}},\ }\href {https://doi.org/10.1103/RevModPhys.90.015002} {\bibfield  {journal} {\bibinfo  {journal} {Reviews of Modern Physics}\ }\textbf {\bibinfo {volume} {90}},\ \bibinfo {pages} {015002} (\bibinfo {year} {2018})}\BibitemShut {NoStop}%
\bibitem [{\citenamefont {Zhou}\ \emph {et~al.}(2020)\citenamefont {Zhou}, \citenamefont {Wang}, \citenamefont {Choi}, \citenamefont {Pichler},\ and\ \citenamefont {Lukin}}]{zho20}%
  \BibitemOpen
  \bibfield  {author} {\bibinfo {author} {\bibfnamefont {L.}~\bibnamefont {Zhou}}, \bibinfo {author} {\bibfnamefont {S.-T.}\ \bibnamefont {Wang}}, \bibinfo {author} {\bibfnamefont {S.}~\bibnamefont {Choi}}, \bibinfo {author} {\bibfnamefont {H.}~\bibnamefont {Pichler}},\ and\ \bibinfo {author} {\bibfnamefont {M.~D.}\ \bibnamefont {Lukin}},\ }\href {https://doi.org/10.1103/PhysRevX.10.021067} {\bibfield  {journal} {\bibinfo  {journal} {Physical Review X}\ }\textbf {\bibinfo {volume} {10}},\ \bibinfo {pages} {021067} (\bibinfo {year} {2020})}\BibitemShut {NoStop}%
\bibitem [{\citenamefont {Farhi}\ \emph {et~al.}()\citenamefont {Farhi}, \citenamefont {Goldstone},\ and\ \citenamefont {Gutmann}}]{far14}%
  \BibitemOpen
  \bibfield  {author} {\bibinfo {author} {\bibfnamefont {E.}~\bibnamefont {Farhi}}, \bibinfo {author} {\bibfnamefont {J.}~\bibnamefont {Goldstone}},\ and\ \bibinfo {author} {\bibfnamefont {S.}~\bibnamefont {Gutmann}},\ }\href {https://doi.org/10.48550/arXiv.1411.4028} {\bibinfo {title} {A {Quantum} {Approximate} {Optimization} {Algorithm}}},\ \bibinfo {note} {arXiv:1411.4028}\BibitemShut {NoStop}%
\bibitem [{\citenamefont {Hadfield}\ \emph {et~al.}(2019)\citenamefont {Hadfield}, \citenamefont {Wang}, \citenamefont {O'Gorman}, \citenamefont {Rieffel}, \citenamefont {Venturelli},\ and\ \citenamefont {Biswas}}]{had19}%
  \BibitemOpen
  \bibfield  {author} {\bibinfo {author} {\bibfnamefont {S.}~\bibnamefont {Hadfield}}, \bibinfo {author} {\bibfnamefont {Z.}~\bibnamefont {Wang}}, \bibinfo {author} {\bibfnamefont {B.}~\bibnamefont {O'Gorman}}, \bibinfo {author} {\bibfnamefont {E.~G.}\ \bibnamefont {Rieffel}}, \bibinfo {author} {\bibfnamefont {D.}~\bibnamefont {Venturelli}},\ and\ \bibinfo {author} {\bibfnamefont {R.}~\bibnamefont {Biswas}},\ }\href {https://doi.org/10.3390/a12020034} {\bibfield  {journal} {\bibinfo  {journal} {Algorithms}\ }\textbf {\bibinfo {volume} {12}},\ \bibinfo {pages} {34} (\bibinfo {year} {2019})}\BibitemShut {NoStop}%
\bibitem [{\citenamefont {Kandala}\ \emph {et~al.}(2017)\citenamefont {Kandala}, \citenamefont {Mezzacapo}, \citenamefont {Temme}, \citenamefont {Takita}, \citenamefont {Brink}, \citenamefont {Chow},\ and\ \citenamefont {Gambetta}}]{kan17}%
  \BibitemOpen
  \bibfield  {author} {\bibinfo {author} {\bibfnamefont {A.}~\bibnamefont {Kandala}}, \bibinfo {author} {\bibfnamefont {A.}~\bibnamefont {Mezzacapo}}, \bibinfo {author} {\bibfnamefont {K.}~\bibnamefont {Temme}}, \bibinfo {author} {\bibfnamefont {M.}~\bibnamefont {Takita}}, \bibinfo {author} {\bibfnamefont {M.}~\bibnamefont {Brink}}, \bibinfo {author} {\bibfnamefont {J.~M.}\ \bibnamefont {Chow}},\ and\ \bibinfo {author} {\bibfnamefont {J.~M.}\ \bibnamefont {Gambetta}},\ }\href {https://doi.org/10.1038/nature23879} {\bibfield  {journal} {\bibinfo  {journal} {Nature}\ }\textbf {\bibinfo {volume} {549}},\ \bibinfo {pages} {242} (\bibinfo {year} {2017})}\BibitemShut {NoStop}%
\bibitem [{\citenamefont {Cerezo}\ \emph {et~al.}(2021)\citenamefont {Cerezo}, \citenamefont {Arrasmith}, \citenamefont {Babbush}, \citenamefont {Benjamin}, \citenamefont {Endo}, \citenamefont {Fujii}, \citenamefont {McClean}, \citenamefont {Mitarai}, \citenamefont {Yuan}, \citenamefont {Cincio},\ and\ \citenamefont {Coles}}]{cer21}%
  \BibitemOpen
  \bibfield  {author} {\bibinfo {author} {\bibfnamefont {M.}~\bibnamefont {Cerezo}}, \bibinfo {author} {\bibfnamefont {A.}~\bibnamefont {Arrasmith}}, \bibinfo {author} {\bibfnamefont {R.}~\bibnamefont {Babbush}}, \bibinfo {author} {\bibfnamefont {S.~C.}\ \bibnamefont {Benjamin}}, \bibinfo {author} {\bibfnamefont {S.}~\bibnamefont {Endo}}, \bibinfo {author} {\bibfnamefont {K.}~\bibnamefont {Fujii}}, \bibinfo {author} {\bibfnamefont {J.~R.}\ \bibnamefont {McClean}}, \bibinfo {author} {\bibfnamefont {K.}~\bibnamefont {Mitarai}}, \bibinfo {author} {\bibfnamefont {X.}~\bibnamefont {Yuan}}, \bibinfo {author} {\bibfnamefont {L.}~\bibnamefont {Cincio}},\ and\ \bibinfo {author} {\bibfnamefont {P.~J.}\ \bibnamefont {Coles}},\ }\href {https://doi.org/10.1038/s42254-021-00348-9} {\bibfield  {journal} {\bibinfo  {journal} {Nature Reviews Physics}\ }\textbf {\bibinfo {volume} {3}},\ \bibinfo {pages} {625} (\bibinfo {year} {2021})}\BibitemShut {NoStop}%
\bibitem [{\citenamefont {Tilly}\ \emph {et~al.}(2022)\citenamefont {Tilly}, \citenamefont {Chen}, \citenamefont {Cao}, \citenamefont {Picozzi}, \citenamefont {Setia}, \citenamefont {Li}, \citenamefont {Grant}, \citenamefont {Wossnig}, \citenamefont {Rungger}, \citenamefont {Booth},\ and\ \citenamefont {Tennyson}}]{til22}%
  \BibitemOpen
  \bibfield  {author} {\bibinfo {author} {\bibfnamefont {J.}~\bibnamefont {Tilly}}, \bibinfo {author} {\bibfnamefont {H.}~\bibnamefont {Chen}}, \bibinfo {author} {\bibfnamefont {S.}~\bibnamefont {Cao}}, \bibinfo {author} {\bibfnamefont {D.}~\bibnamefont {Picozzi}}, \bibinfo {author} {\bibfnamefont {K.}~\bibnamefont {Setia}}, \bibinfo {author} {\bibfnamefont {Y.}~\bibnamefont {Li}}, \bibinfo {author} {\bibfnamefont {E.}~\bibnamefont {Grant}}, \bibinfo {author} {\bibfnamefont {L.}~\bibnamefont {Wossnig}}, \bibinfo {author} {\bibfnamefont {I.}~\bibnamefont {Rungger}}, \bibinfo {author} {\bibfnamefont {G.~H.}\ \bibnamefont {Booth}},\ and\ \bibinfo {author} {\bibfnamefont {J.}~\bibnamefont {Tennyson}},\ }\href {https://doi.org/10.1016/j.physrep.2022.08.003} {\bibfield  {journal} {\bibinfo  {journal} {Physics Reports}\ }\textbf {\bibinfo {volume} {986}},\ \bibinfo {pages} {1} (\bibinfo {year} {2022})}\BibitemShut {NoStop}%
\bibitem [{\citenamefont {Verstraete}\ \emph {et~al.}(2009)\citenamefont {Verstraete}, \citenamefont {Wolf},\ and\ \citenamefont {Ignacio~Cirac}}]{ver09}%
  \BibitemOpen
  \bibfield  {author} {\bibinfo {author} {\bibfnamefont {F.}~\bibnamefont {Verstraete}}, \bibinfo {author} {\bibfnamefont {M.~M.}\ \bibnamefont {Wolf}},\ and\ \bibinfo {author} {\bibfnamefont {J.}~\bibnamefont {Ignacio~Cirac}},\ }\href {https://doi.org/10.1038/nphys1342} {\bibfield  {journal} {\bibinfo  {journal} {Nature Physics}\ }\textbf {\bibinfo {volume} {5}},\ \bibinfo {pages} {633} (\bibinfo {year} {2009})}\BibitemShut {NoStop}%
\bibitem [{\citenamefont {Cho}\ \emph {et~al.}(2011)\citenamefont {Cho}, \citenamefont {Bose},\ and\ \citenamefont {Kim}}]{cho11}%
  \BibitemOpen
  \bibfield  {author} {\bibinfo {author} {\bibfnamefont {J.}~\bibnamefont {Cho}}, \bibinfo {author} {\bibfnamefont {S.}~\bibnamefont {Bose}},\ and\ \bibinfo {author} {\bibfnamefont {M.~S.}\ \bibnamefont {Kim}},\ }\href {https://doi.org/10.1103/PhysRevLett.106.020504} {\bibfield  {journal} {\bibinfo  {journal} {Physical Review Letters}\ }\textbf {\bibinfo {volume} {106}},\ \bibinfo {pages} {020504} (\bibinfo {year} {2011})}\BibitemShut {NoStop}%
\bibitem [{\citenamefont {Cubitt}(2023)}]{cub23}%
  \BibitemOpen
  \bibfield  {author} {\bibinfo {author} {\bibfnamefont {T.~S.}\ \bibnamefont {Cubitt}},\ }\href {https://doi.org/10.48550/arXiv.2303.11962} {\bibinfo {title} {Dissipative ground state preparation and the {Dissipative} {Quantum} {Eigensolver}}} (\bibinfo {year} {2023}),\ \bibinfo {note} {arXiv:2303.11962}\BibitemShut {NoStop}%
\bibitem [{\citenamefont {Nielsen}\ and\ \citenamefont {Chuang}(2011)}]{nie11}%
  \BibitemOpen
  \bibfield  {author} {\bibinfo {author} {\bibfnamefont {M.~A.}\ \bibnamefont {Nielsen}}\ and\ \bibinfo {author} {\bibfnamefont {I.~L.}\ \bibnamefont {Chuang}},\ }\href@noop {} {\emph {\bibinfo {title} {Quantum {Computation} and {Quantum} {Information}: 10th {Anniversary} {Edition}}}}\ (\bibinfo  {publisher} {Cambridge University Press},\ \bibinfo {address} {Cambridge; New York},\ \bibinfo {year} {2011})\BibitemShut {NoStop}%
\bibitem [{\citenamefont {Cho}\ \emph {et~al.}(2015)\citenamefont {Cho}, \citenamefont {Bose},\ and\ \citenamefont {Kim}}]{cho15}%
  \BibitemOpen
  \bibfield  {author} {\bibinfo {author} {\bibfnamefont {J.}~\bibnamefont {Cho}}, \bibinfo {author} {\bibfnamefont {S.}~\bibnamefont {Bose}},\ and\ \bibinfo {author} {\bibfnamefont {M.~S.}\ \bibnamefont {Kim}},\ }\href {https://doi.org/10.1016/j.optcom.2014.06.059} {\bibfield  {journal} {\bibinfo  {journal} {Optics Communications}\ }\textbf {\bibinfo {volume} {337}},\ \bibinfo {pages} {66} (\bibinfo {year} {2015})}\BibitemShut {NoStop}%
\bibitem [{\citenamefont {Amin}(2008)}]{ami08}%
  \BibitemOpen
  \bibfield  {author} {\bibinfo {author} {\bibfnamefont {M.~H.~S.}\ \bibnamefont {Amin}},\ }\href {https://doi.org/10.1103/PhysRevLett.100.130503} {\bibfield  {journal} {\bibinfo  {journal} {Physical Review Letters}\ }\textbf {\bibinfo {volume} {100}},\ \bibinfo {pages} {130503} (\bibinfo {year} {2008})}\BibitemShut {NoStop}%
\bibitem [{\citenamefont {Werner}\ \emph {et~al.}(2023)\citenamefont {Werner}, \citenamefont {Garc\'ia-S\'aez},\ and\ \citenamefont {Estarellas}}]{wer23}%
  \BibitemOpen
  \bibfield  {author} {\bibinfo {author} {\bibfnamefont {M.}~\bibnamefont {Werner}}, \bibinfo {author} {\bibfnamefont {A.}~\bibnamefont {Garc\'ia-S\'aez}},\ and\ \bibinfo {author} {\bibfnamefont {M.~P.}\ \bibnamefont {Estarellas}},\ }\href {https://doi.org/10.1103/PhysRevResearch.5.043236} {\bibfield  {journal} {\bibinfo  {journal} {Physical Review Research}\ }\textbf {\bibinfo {volume} {5}},\ \bibinfo {pages} {043236} (\bibinfo {year} {2023})}\BibitemShut {NoStop}%
\bibitem [{\citenamefont {Eschner}\ \emph {et~al.}(2003)\citenamefont {Eschner}, \citenamefont {Morigi}, \citenamefont {Schmidt-Kaler},\ and\ \citenamefont {Blatt}}]{esc03}%
  \BibitemOpen
  \bibfield  {author} {\bibinfo {author} {\bibfnamefont {J.}~\bibnamefont {Eschner}}, \bibinfo {author} {\bibfnamefont {G.}~\bibnamefont {Morigi}}, \bibinfo {author} {\bibfnamefont {F.}~\bibnamefont {Schmidt-Kaler}},\ and\ \bibinfo {author} {\bibfnamefont {R.}~\bibnamefont {Blatt}},\ }\href@noop {} {\bibfield  {journal} {\bibinfo  {journal} {Journal of the Optical Society of America B}\ }\textbf {\bibinfo {volume} {20}},\ \bibinfo {pages} {1003} (\bibinfo {year} {2003})}\BibitemShut {NoStop}%
\bibitem [{\citenamefont {Grover}(1996)}]{gro96}%
  \BibitemOpen
  \bibfield  {author} {\bibinfo {author} {\bibfnamefont {L.~K.}\ \bibnamefont {Grover}},\ }in\ \href {https://doi.org/10.1145/237814.237866} {\emph {\bibinfo {booktitle} {Proceedings of the twenty-eighth annual {ACM} symposium on {Theory} of {Computing}}}},\ \bibinfo {series and number} {{STOC} '96}\ (\bibinfo  {publisher} {Association for Computing Machinery},\ \bibinfo {address} {New York, NY, USA},\ \bibinfo {year} {1996})\ pp.\ \bibinfo {pages} {212--219}\BibitemShut {NoStop}%
\bibitem [{\citenamefont {Bennett}\ \emph {et~al.}(1997)\citenamefont {Bennett}, \citenamefont {Bernstein}, \citenamefont {Brassard},\ and\ \citenamefont {Vazirani}}]{ben97}%
  \BibitemOpen
  \bibfield  {author} {\bibinfo {author} {\bibfnamefont {C.~H.}\ \bibnamefont {Bennett}}, \bibinfo {author} {\bibfnamefont {E.}~\bibnamefont {Bernstein}}, \bibinfo {author} {\bibfnamefont {G.}~\bibnamefont {Brassard}},\ and\ \bibinfo {author} {\bibfnamefont {U.}~\bibnamefont {Vazirani}},\ }\href {https://doi.org/10.1137/S0097539796300933} {\bibfield  {journal} {\bibinfo  {journal} {SIAM Journal on Computing}\ }\textbf {\bibinfo {volume} {26}},\ \bibinfo {pages} {1510} (\bibinfo {year} {1997})}\BibitemShut {NoStop}%
\bibitem [{\citenamefont {Boyer}\ \emph {et~al.}(1998)\citenamefont {Boyer}, \citenamefont {Brassard}, \citenamefont {H\o{}yer},\ and\ \citenamefont {Tapp}}]{boy98}%
  \BibitemOpen
  \bibfield  {author} {\bibinfo {author} {\bibfnamefont {M.}~\bibnamefont {Boyer}}, \bibinfo {author} {\bibfnamefont {G.}~\bibnamefont {Brassard}}, \bibinfo {author} {\bibfnamefont {P.}~\bibnamefont {H\o{}yer}},\ and\ \bibinfo {author} {\bibfnamefont {A.}~\bibnamefont {Tapp}},\ }\href {https://doi.org/10.1002/(SICI)1521-3978(199806)46:4/5<493::AID-PROP493>3.0.CO;2-P} {\bibfield  {journal} {\bibinfo  {journal} {Fortschritte der Physik}\ }\textbf {\bibinfo {volume} {46}},\ \bibinfo {pages} {493} (\bibinfo {year} {1998})}\BibitemShut {NoStop}%
\bibitem [{\citenamefont {Zalka}(1999)}]{zal99}%
  \BibitemOpen
  \bibfield  {author} {\bibinfo {author} {\bibfnamefont {C.}~\bibnamefont {Zalka}},\ }\href {https://doi.org/10.1103/PhysRevA.60.2746} {\bibfield  {journal} {\bibinfo  {journal} {Physical Review A}\ }\textbf {\bibinfo {volume} {60}},\ \bibinfo {pages} {2746} (\bibinfo {year} {1999})}\BibitemShut {NoStop}%
\bibitem [{\citenamefont {Shor}(1994)}]{sho94}%
  \BibitemOpen
  \bibfield  {author} {\bibinfo {author} {\bibfnamefont {P.}~\bibnamefont {Shor}},\ }in\ \href {https://doi.org/10.1109/SFCS.1994.365700} {\emph {\bibinfo {booktitle} {Proceedings 35th {Annual} {Symposium} on {Foundations} of {Computer} {Science}}}}\ (\bibinfo {year} {1994})\ pp.\ \bibinfo {pages} {124--134}\BibitemShut {NoStop}%
\end{thebibliography}


\end{document}